\documentclass[twocolumn,showpacs,preprintnumbers,aps,amssymb,amsmath,secnumarabic,floatfix, superscriptaddress,prl]{revtex4}

\usepackage{graphicx}
\usepackage{epsfig}
\usepackage{float}
\usepackage{amsmath} 
\usepackage{amsfonts} 
\usepackage{amssymb}

\newcommand{\etal}{{\it et~al}}

\begin{document}
\title{Zero vector potential mechanism of attosecond absorption in strongly relativistic
       plasmas}
\author{T.~Baeva}
\affiliation{Central Laser Facility, Rutherford-Appleton Laboratory, Chilton, Oxon., OX11 0QX, UK}
\author{S.~Gordienko}
\affiliation{Landau Institute for Theoretical Physics, Kosygina 2, 119334 Moscow, Russia}
\author{A.P.L.~Robinson}
\affiliation{Central Laser Facility, Rutherford-Appleton Laboratory, Chilton, Oxon., OX11 0QX, UK}
\author{P.A.~Norreys}
\affiliation{Central Laser Facility, Rutherford-Appleton Laboratory, Chilton, Oxon., OX11 0QX, UK}

\maketitle

{\bf 
The understanding of the physics of laser-matter interactions in the strongly relativistic regime is of fundamental importance. In this article, a new mechanism of fast electron generation at the vacuum-solid boundary of intense laser pulse interaction with overdense plasma is described. It is one that has no analogue in classical, non-relativistic laser-plasma interactions. Here, conclusive proof is provided that the key contribution to the fast electron generation is given by the zero points of the vector potential. We demonstrate that the new mechanism leads to scalings for the fast electron energy, which explicitly depend on the plasma density, thus providing a new insight into relativistic laser-matter interaction. Furthermore, it is shown that this new mechanism provides the dominant contribution to the interaction by the injection of energy into the overdense plasma delivered by attosecond-duration electron bunches. This new understanding will allow the future generation of a single ultra-bright attosecond  X-ray pulse by suitable control of the laser pulse polarization \cite{Gordienko2004}-\cite{Nomura}. This process will also allow single pulse attosecond electron bunches to be generated that can be further accelerated in laser wakefield accelerators \cite{Mangles}-\cite{Faure}. Other applications that will benefit from this new insight include laser driven ion accelerators \cite{Fuchs}-\cite{Robinson2008}, fast ignition inertial confinement fusion \cite{Tabak}-\cite{Naumova} as well as fundamental studies at the intensity frontier \cite{Mourou}-\cite{Norreys}.
}

Possible mechanisms for generation of fast electrons have been extensively investigated in recent years. A well known and widely quoted scaling for the energy of the fast electrons was proposed by Wilks \cite{Wilks}. This scaling was observed in several experiments (e.g. \cite{Malka} and \cite{McKenna}), however, other experimental and numerical works present a more complex picture (e.g. \cite{Santos}). 
In terms of the absorption fraction, a recent review of many experimental and theoretical studies of relativistic laser absorption \cite{DaviesAbs} showed that absorption fractions from a few percent to nearly 100 percent have been measured or calculated by numerical simulations. This suggests that there are several regimes of absorption. Despite the absence of agreement in the laser-plasma 
community on the best scaling or theoretical description of absorption, there is a consensus that the ponderomotive {\bf jxB}-force plays an important role in generating relativistic electrons.

The mechanism described in this article is, by nature, a non-ponderomotive mechanism of absorption during interaction of ultra-intense laser pulses with sharp density gradients. This theory predicts the generation of fast electrons propagating into the plasma bulk and coherent x-rays in the reflected from the plasma radiation in a single self-consistent framework. The most peculiar and novel feature of this theory is that the generation of bunches of high energy electrons depends on the existence of zeroes in the vector potential and not 
on fluctuations in the incident laser intensity alone. For this reason we refer to this new mechanism as the Zero Vector Potential mechanism. All previously proposed mechanisms - including {\bf jxB}-heating \cite{KruerEstabrook}, vacuum heating \cite{Gibbon}, Brunel heating \cite{Brunel}, anharmonic resonance \cite{Mulser} etc - make no such claim. This Article describes the new absorption mechanism, demonstrates the new scalings for the fast electron energy and presents numerical simulations which demonstrate the predictions of the zero vector potential theory for efficient control of the energy absorption.

The existence of at least two regimes of absorption during intense laser-plasma interaction can be understood on the basis of simple physical reasoning. If we denote the frequency of the incident laser radiation as $\omega$ then the {\bf jxB}-force oscillates with frequency 2$\omega$, which is much lower than the frequency of relativistic electron oscillations in the overdense plasma ($\omega_p=\sqrt{4\pi n_e e^2/a_0m_e}$, where $n_e$ is the plasma density). Consequently, the question arises, in which density region does the main interaction between the incident radiation and the target take place? It is clear that if the region where the laser frequency is of the order of the frequency of relativistic electron oscillations is broad enough, then the interaction is localised in this region and the {\bf jxB}-force indeed drives the electrons to high energies, which they then transport inside the plasma, since plasma effects are not fast enough to counteract this ponderomotive force. 
  
\begin{figure} [h]
\centerline{\includegraphics[width=5.5cm,clip]{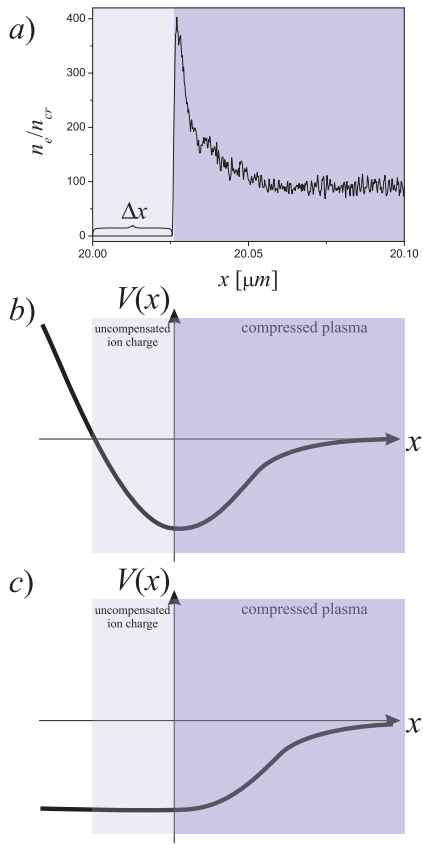}}
\caption{a) For plasmas of steep density gradient the pressure of the incident radiation can significantly compress the plasma thus shifting the interaction area to high plasma densities as particle-in-cell simulations \cite{Robinson2008},\cite{Robinson2007},\cite{Robinson2009} clearly demonstrate. b) During the adiabatic phase the laser light pressure and the electrostatic Coulomb force are in equilibrium building a potential well in which the electron oscillations are confined; c) As the vector potential goes through zero the adiabatic equilibrium is disrupted allowing fast electrons to leave the potential
well.}
\label{densityPeak}
\end{figure}

However, a completely different regime takes place if the density gradient in the vicinity of the critical density is steep. In that case, the pressure of the incident radiation can significantly compress the plasma and shift the interaction area to plasma densities which are much higher than the relativistic critical density (Fig.~\ref{densityPeak} a). As a result, in the case of a steep density gradient, the laser-plasma interaction takes place at plasma densities well above the relativistic critical plasma density. If this is the case, the frequency of the {\bf jxB}-force is much lower than the frequency of relativistic electron oscillations at the area determining the energy transfer from the incident laser pulse to the plasma. Therefore the plasma is able to respond fast enough to compensate the ponderomotive force by the Coulomb attraction between ions and electrons. In other words, the plasma adjusts its state adiabatically to this force.

Consequently we are in an adiabatic regime in which electrons are not directly accelerated by the {\bf jxB}-force. Instead, the skin layer is in equilibrium between the ponderomotive pressure and the electrostatic Coulomb force. These two forces create a potential well in which the electrons are confined to perform finite oscillations. The potential well can be visualised if we consider the Hamiltonian of a single electron in the plasma skin layer under the action of the field of the laser pulse. Due to conservation of momentum in direction tangential to the plasma we have

\begin{equation} \label{pA}
  p_{\tau} = \frac{eA_{\tau}}{c},
\end{equation}

\noindent
where $p_{\tau}$ is the tangential electron and $A_{\tau}$ is the tangential vector potential. The Hamiltonian of a single electron can now be written as
  
\begin{equation} \label{hamiltonian}
  H = \sqrt{m_e^2c^2 + p_x^2 + e^2A_{\tau}^2/c^2} + \Phi
\end{equation}
  
\noindent
revealing that the potential well in which the electrons are confined when the laser light pressure and the electrostatic Coulomb force are in equilibrium consists of the electrostatic part $\Phi$ vanishing inside the quasi-neutral plasma and the linear part connected with the vector potential $A_{\tau}$ outside the plasma slab (Fig.~\ref{densityPeak} b).

The equilibrium is disrupted for short times when the vector potential goes through zero. Since at these moments the ponderomotive pressure vanishes, one of the walls of the potential well is suppressed (Fig.~\ref{densityPeak} c), causing a violent reconstruction of the plasma skin layer.

This reconstruction of the plasma skin layer has two important consequences. Firstly, the electrostatic Coulomb energy, stored in the skin layer over the adiabatic phase, is freed during the reconstruction in the form of fast electrons leaving the plasma in the specular direction and the violent electron dynamics leads to bursts of coherent 
x-ray radiation in vacuum in the same direction. Secondly, upon returning the fast electrons enter the plasma, thus transporting energy irreversibly into the target.

It is interesting to note that the process of reconstruction can be understood in rather simple terms, owing to a fascinating phenomenon which occurs only in relativistic skin-layers. Indeed, it turns out that in the region which contributes to the interaction of the overdense plasma with the relativistic radiation the zero point of the vector potential moves with velocity $c$ \cite{Baevabook}. In order to understand this phenomenon let us consider closely the relativistic overdense plasma slab and a laser electromagnetic field normally incident onto the plasma for the sake of simplicity. Since the slab is overdense it expels the electromagnetic field and since the electrons are ultra-relativistic the electromagnetic field is expelled with the speed of light. On the other hand the transverse motion of the electrons is connected to the vector potential according to Eq.~(\ref{pA}). The longitudinal velocity of the electrons is related to the vector potential by

\begin{equation} \label{vx}
  v_x = \frac{p_xc}{\sqrt{m_e^2c^2 + p_x^2 + e^2A_{\tau}^2/c^2}}.
\end{equation}

\noindent
This means that at the zero of the vector potential the electron momentum $p$ vanishes and the relativistic electrons which are in the vicinity of this point move with velocity $c$ either in the direction out of or into the plasma. Thus the ultra-relativistic electrons 
stick to the zero of the vector potential, leading to the zero being "dressed" by co-moving relativistic electrons during its motion through the plasma. In its turn this overdense slab of ultra-relativistic electrons expels the electromagnetic field with velocity $c$, thus 
closing the self-consistent picture of the motion of the vector potential zero.  No confusion has arisen here between formulating the ponderomotive force in either the ${\bf E},{\bf B}$ or $\phi,{\bf A}$ formulation. One arrives at the same conclusions with either formulation.

Note that the longer the distance that the zero travels in the skin layer, the higher the number of electrons that stick to this zero. The momentum of the trapped electron bunch increases due to the electrostatic attraction to the ions until the electrons leave the ion 
background.  After that the Coulomb attraction to the immobile ions starts decelerating the electrons to return them back into the plasma. These returning electrons have already consumed the electrostatic energy of the skin layer. As a result they form a short electron 
pulse of relativistic electrons injecting energy into the plasma slab.
  
It must be emphasized that it is this motion of the zero in the vector potential that defines this mechanism as having no non-relativistic classical analogue.  The reason for this is that only in a relativistic plasma can one have a vector potential with a moving zero in the skin layer. In a non-relativistic plasma, the vector potential can only exponentially decay in the skin layer. Thus the zero vector potential mechanism is a unique product of relativistic electrodynamics.

Note that due to their relativistic motion the coherently moving electrons attached to the vector potential zero radiate in phase, leading to the generation of a large number of coherent x-rays during the laser-plasma interaction. These coherent x-rays were observed in a 
number of numerical simulations \cite{Gordienko2004} and experiments \cite{Dromey2006,Dromey2007}. It was shown analytically by the theory of relativistic spikes that the spectrum of these x-rays is universal and obeys a power-law decay, where the intensity of the frequency $\omega$ scales as \cite{Baeva2006, Baeva2006rpc}

\begin{equation} \label{universalSpectrum}
  I(\omega) \propto \left(\frac{\omega}{\omega_0}\right)^{-8/3}
\end{equation}

\noindent
up to a roll-off frequency on the order of $\omega_{roll} \propto I_0^{-3/2}$, in agreement with experimental observations.

From these arguments, one can derive a scaling relation for the energy of the fast electrons.  In what follows we will pursue a simple derivation for the case of a step-like interface.  This can be derived far more rigorously using relativistic similarity theory \cite{Baevabook, Gordienko2005}.
  
In the adiabatic phase of the interaction the electrostatic response of the plasma balances the radiation pressure.  Noting this, let us estimate the distance ${\Delta}x$ by which the electron fluid is displaced into the plasma (see Fig.~\ref{densityPeak} a). When the electron fluid is pushed into the plasma a non-compensated ion charge $Q=en_e{\Delta}x$ is left at the plasma boundary. The attractive force acting on the electron fluid from the side of the ions is therefore
$F_i \propto e^2n_e^2({\Delta}x)^2$. In the adiabatic phase, this force is compensated by the force acting on the electrons from the side of the laser light pressure
$F_p=\frac{E_{laser}^2}{2\pi}\propto\left(a_0^2\omega_0^2/e^2\right)\left(m_ec\right)^2$,
thus leading to the following estimation for the electron fluid displacement

\begin{equation}
  {\Delta}x=\frac{\lambda}{S}.
\end{equation}

\noindent Here $\lambda$ is the laser wavelength and $S$ is a dimensionless combination of the plasma density $n_e$, the critical density $n_{cr}$ and the dimensionless vector potential $a_0$, such that $S=n_e/n_{cr}a_0$ \cite{Gordienko2005}.
 
The above estimation essentially defines how much energy is stored in the `plasma capacitor' during the adiabatic phase. During the $A_{\tau}=0$ phase the electrons discharge the capacitor and get the energy stored in it. Thus the energy of the fast electrons can be estimated as the energy which they gain when crossing the capacitor of width $\Delta x$:

\begin{equation} \label{FEenergy}
  W=eE_{laser}{\Delta}x
    \propto e^2n_e({\Delta}x)^2\propto\frac{a_0}{S}(m_ec^2).
\end{equation}

\noindent
Eq.~(\ref{FEenergy}) gives important insight into the absorption of energy in relativistic laser-matter interaction. It shows that when the interaction takes place at the critical surface ($n_e = a_0n_{cr}$, corresponding to $S\approx 1$), the fast electron energy is proportional to $a_0$. This corresponds to the ponderomotive regime discussed above, in which Eq.~(~\ref{FEenergy}) reproduces the Wilks' scaling. In the adiabatic regime however, the interaction takes place at much higher densities ($S\gg 1$) leading to
    
\begin{equation} \label{newScaling}
  W\propto\frac{a_0^2n_{cr}}{n_e}(m_ec^2).
\end{equation}

\begin{figure} [h]
    \centerline{\includegraphics[width=6.5cm,clip]{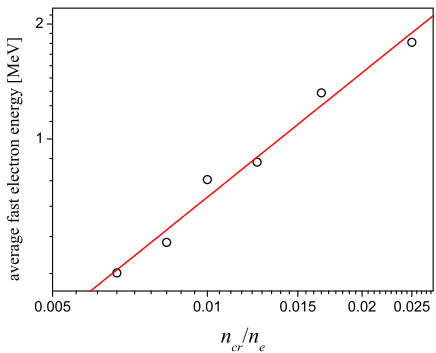}}
    \caption{Log-log plot of the inverse-linear dependence of the fast electron energy 
             from the plasma density (Eq.~(\ref{newScaling})), observed in PIC simulations 
             for a wide range of densities.}
    \label{scalingSimulation}
  \end{figure}

\noindent
This explicit dependence of the fast electron energy on the plasma density underlines the important difference between the zero vector potential mechanism and the ponderomotive ${\bf jxB}$-mechanism. The density dependence predicted by the new zero vector potential mechanism
can be demonstrated by a PIC simulation. Fig.~(\ref{scalingSimulation}) shows that the inverse-linear dependence of the fast electron energy on the density can be observed for a broad range of plasma densities, thus confirming the predictions of the new absorption mechanism.

We can also estimate the number of fast electrons generated, as the number of non-compensated ion charges in the plasma capacitor $N=n_e{\Delta}x\propto a_0\lambda n_{cr}$.
Note that the number of fast electrons generated by this process does not depend on the plasma density, but only on the intensity and frequency of the laser.

The physical arguments presented above suggest that the zero points of the vector potential are responsible for the generation of fast electrons in relativistic laser-overdense plasma interaction and thus for the absorption of energy in these plasmas. In what follows we apply numerical simulations in order to demonstrate how the production of fast electrons and coherent x-rays is governed by the zero points of the vector potential. For this purpose we consider the interaction of a steep plasma slab with relativistically intense laser pulses of different polarization thus varying the number of zero points of the vector potential while keeping the oscillation behaviour of the 
ponderomotive pressure unchanged.

We first consider the interaction of two laser pulses of linear and elliptical polarization with the plasma slab of density $n_e=80n_{cr}$ under normal incidence using a 1D3P particle-in-cell code \cite{Robinson2008},\cite{Robinson2007},\cite{Robinson2009}. The wavelength of both pulses is $\lambda$=1$\mu$m and their duration is $\tau$=5fs. The vector potential for the linear polarization propagates in 
vacuum as a plain wave
  
\begin{equation}
  a_y=a_{y0}\exp\left(-(x-ct)^2/2c^2\tau^2\right)
        \cos\left(2\pi(x-ct)/\lambda\right), \nonumber
\end{equation}
  
\noindent
where $a_{y0}=20$, while the components of the laser pulse with elliptical polarization are 
  
\begin{eqnarray}
    a_y&=&a_{y0}\exp\left(-(x-ct)^2/2c^2\tau^2\right)
          \cos\left(2\pi(x-ct)/\lambda\right) \nonumber\\
    a_z&=&a_{z0}\exp\left(-(x-ct)^2/2c^2\tau^2\right)
          \sin\left(2\pi(x-ct)/\lambda\right), \nonumber
\end{eqnarray}
  
\noindent
where $a_{y0}=\sqrt{425}$ and $a_{z0}=5$. Note that this choice of amplitudes insures that for the laser light pressure given by
$P\propto a_y^2+a_z^2\propto\left(a_{y0}^2-a_{z0}^2\right)\cos^2  
(\omega t)+a_{z0}^2$
  the amplitude of laser pressure oscillations is the same for both polarizations. Note that 
  although the light pressure oscillations are identical, the linearly polarized pulse has 
  several vector potential zeroes, while the elliptical pulse has no zero crossings.
  
  \begin{figure} [h]
    \centerline{\includegraphics[width=8.5cm,clip]{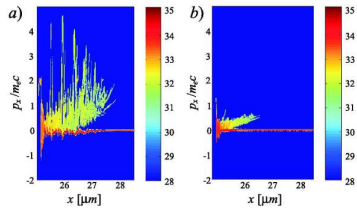}}
    \caption{Logarithmic colour plot snapshots of electron plasma density as a 
             function of longitudinal electron momentum $p_x$ and coordinate $x$ for 
             the case of a) linear polarization and b) elliptical polarization. 
             Although the amplitude of laser pressure oscillations is the same in 
             both cases the fast electron shocks disappear in the elliptical case 
             due to the lack of zeroes in the vector potential. Colour scale is in 
             units of kgm$^2$s$^{-1}$.}
    \label{pxx_linellipt}
  \end{figure}
  
  \begin{figure} [h]
    \centerline{\includegraphics[width=8cm,clip]{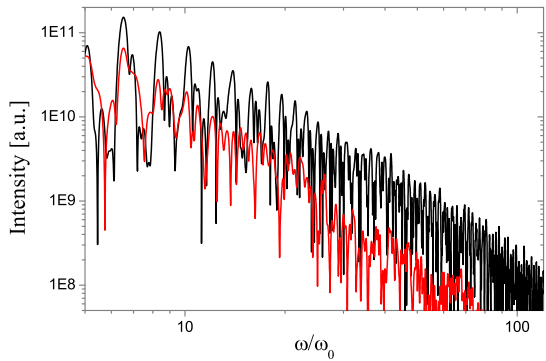}}
    \caption{Spectra of coherent x-ray radiation in the signal reflected from the plasma for
             linear (black line) and elliptical polarization (red line). Due to the absence
             of fast electrons in the case of elliptical polarization 
             (compare Fig.~\ref{pxx_linellipt} b) the production of coherent x-rays is
             significantly reduced.}
    \label{spectra_linellipt}
  \end{figure}
  
  Fig.~\ref{pxx_linellipt} shows the electron distribution function in colour as a function 
  of the electron momentum in longitudinal direction $p_x$ and the longitudinal coordinate 
  $x$. One can clearly see that during the interaction of the plasma slab with the laser pulse 
  of linear polarization electron shocks are generated in the skin layer and propagate into the 
  bulk of the plasma slab twice per laser period (distance $\lambda$/2 apart). At the head of 
  the plasma one can clearly observe electrons with positive and negative momenta of equal size,
  representing the oscillations of the electrons in the potential well produced by the laser 
  vector potential and the electrostatic Coulomb force.

  Although the oscillations of the ponderomotive potential have not changed, when we move to   
  the interaction of the plasma slab with the elliptically polarized pulse the picture changes  
  completely. Since there are no zero points of the vector potential, no electron shocks are   
  generated in the plasma skin layer and hence no shocks propagate into the bulk of the plasma.   The elimination of the electron shocks in the simulation with the elliptically polarized 
  pulse cannot be explained by the ${\bf jxB}$-mechanism.
  Note that one can clearly see a flux of low energy electrons leaking from the imperfections 
  of the potential well. Although both snapshots are taken at the same moment of time
  the leaking electrons have lower velocity and thus cannot penetrate into the plasma bulk as 
  deep as the electron shocks.  
  
  The importance of these two simulations can be summarized as follows : We are comparing two cases.  One where the laser
  pulse has a strongly oscillating ponderomotive force and zeroes in the vector potential, and one where we have the same amplitude of
  oscillation in the ponderomotive force, but {\em no} zeroes in the vector potential.  We see that the highly energetic electron shocks are {\em only} 
  produced when we have zeroes in the vector potential.  This numerically demonstrates the essence of the zero vector potential mechanism, and this cannot be 
  explained by the `jxB' mechanism.
  
  \begin{figure} [ht]
    \centerline{\includegraphics[width=7.5cm,clip]{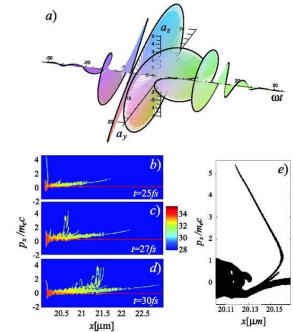}}
    \caption{a) Vector potential of mixed polarization allowing only one zero point;
             b)-d)Logarithmic colour plot of single fast electron bunch generated by a 
             single zero of the vector potential a). b) Generation of a sharp electron 
             bunch inside the plasma skin layer; c) and d) propagation of the bunch into 
             the plasma bulk. The relativistic electron bunch experiences instability during
             the propagation into the cold bulk plasma. e) Zoom of the single electron bunch 
             in b) shows that electrons of 2MeV are confined within a time interval of 5as, 
             and electrons between 0.5MeV and 2.5Mev are propagating within 130as. Colour 
             scale is in units of kgm$^2$s$^{-1}$.}
    \label{vectorPotential_rpc}
  \end{figure}

  Note that according to the physical picture presented above, the electron shocks are the 
  consequence of electrons which were able to escape the potential well during the times the 
  vector potential goes through zero. Before entering the plasma and propagating into the bulk 
  in the form of electron shocks these electrons first leave the plasma skin layer and radiate 
  coherent x-rays in specular direction. Fig.~\ref{spectra_linellipt} shows the spectrum of 
  the reflected electromagnetic field. For the case of linear polarisation this spectrum
  follows the harmonic decay law (\ref{universalSpectrum}) derived in \cite{Baeva2006}. 
  However the x-ray intensity drops dramatically in the case of elliptical polarization, 
  as a result of the missing fast electrons.

  The experience of controlling the generation of electron shocks by means of changing the   
  number of zero points in the vector potential suggests an exciting application of this 
  mechanism of control of the laser plasma interaction. Indeed, the zero vector potential 
  mechanism suggests that if the vector potential of the pulse vanishes only once, an isolated 
  electron bunch can be generated. 
  
  In order to demonstrate this numerically we consider the 
  interaction of the plasma slab with a laser pulse of variable polarization, constructed in 
  such a way as to assure that the vector potential goes through zero only once. The vector 
  potential components of this pulse depend on time as
  
  \begin{eqnarray}
    a_y&=&a_{y0}\exp\left(-(x-ct)^2/2c^2\tau^2\right)
        \cos\left(2\pi(x-ct)/\lambda\right) \nonumber\\
    a_z&=&a_{z0}\exp\left(-(x-ct)^2/2c^2\tau^2\right)
        \cos\left(2\pi(x-ct)/\lambda_c+\pi/8\right),\nonumber
  \end{eqnarray}
  
  \noindent
  where $\lambda_c=800nm$. The vector potential is presented in 
  Fig.~\ref{vectorPotential_rpc} a). 
  
  
  As a result of the interaction of this pulse with the plasma slab a single electron bunch 
  is generated in the plasma skin layer (Fig.~\ref{vectorPotential_rpc} b) and propagates 
  into the plasma bulk (Fig.~\ref{vectorPotential_rpc} c). Note that the electron bunch is
  moving with ultra-relativistic velocity through the bulk of cold plasma, leading to
  instability of the bunch structure (Fig.~\ref{vectorPotential_rpc} d).
  Fig.~\ref{vectorPotential_rpc} e) zooms the isolated electron shock presented in
  Fig.~\ref{vectorPotential_rpc} b) and reveals the spatial width and
  the geometry of this shock. One can observe that electrons of different energies are
  localized in an astonishingly narrow area, thus all electrons of 2MeV energy are inside 
  a 15\AA area (5as duration). This width is much shorter than both the plasma length and 
  the Debye length corresponding to the temperature used in the simulation. This is a 
  direct indication to an unusual mechanism of electron bunch generation which cannot 
  be reduced to standard
  plasma physics phenomena, taking place on longer scalelengths. In the shock observed the
  lower energy electrons run ahead of the higher energy electrons and that electron 
  separation results in a locally monochromatic spectrum of the fast electrons, as 
  explained by the zero vector potential mechanism. Note that when the vector potential 
  zero accompanied by fast relativistic electrons leaves the plasma there is no spatial
  separation over energies, and relativistic electrons of all energies propagate together 
  at about the speed of light. However, the Coulomb attraction to the ions turns back the
  velocities of the electrons with lower energies earlier than of those with higher energies.
  As a result, in the electron shock returning to the plasma, the lower energy electrons run
  ahead. In the simulation presented on Fig.~\ref{vectorPotential_rpc} e) the distance between 
  0.5MeV and 2.5Mev electrons is about 130as.  The generation of a single electron bunch 
  by such an engineered pulse cannot be explained by the ${\bf jxB}$-mechanism, nor can 
  the ${\bf jxB}$-mechanism explain the incredibly short duration of the electron bunches.  

  \begin{figure} [h]
    \centerline{\includegraphics[width=6.5cm,clip]{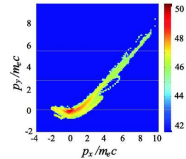}}
    \caption{Logarithmic colour plot of $p_x-p_y$ phase space from boosted 1d3v PIC simulation.
             Colour scale is in units of kgm$^2$s$^{-1}$.}
    \label{oblique}
  \end{figure} 
  
Finally let us emphasize that the zero vector potential mechanism applies not only for the case of normal but also for oblique laser incidence onto the plasma slab. In this case the vector potential zero propagates out of the plasma in specular direction leading to a burst of fast electrons and
coherent x-rays in the same direction. Upon their return to the plasma the fast electrons propagate inwards in the direction of laser pulse propagation. This behaviour can be clearly seen in a particle-in-cell simulation of oblique laser-plasma interaction (Fig.~\ref{oblique}).  If one considers the same interaction from the {\bf jxB} perspective then one would argue that the electron bunches must enter normal to the plasma surface (the direction of the gradient in the ponderomotive potential). Only the zero vector potential mechanism explains why they propagate along the direction of laser pulse propagation.
  
In conclusion, we presented a novel mechanism of energy absorption in relativistic overdense plasmas and derived a new scaling for the energy of fast electrons which explicitly shows dependence on the plasma density. The new non-ponderomotive mechanism explains the relation of relativistic absorption to attosecond time scales. We demonstrated that the generation of fast electrons and coherent x-rays in these plasmas is controlled by the zero points of the vector potential allowing the generation of sharp shocks of fast electrons propagating into the plasma bulk twice per laser period. Controlling the laser pulse polarization in such a way that it goes through zero only once leads to the generation of an isolated fast electron bunch.
    

\end{document}